\documentstyle[11pt,moriond,epsfig]{article}

\def\be{\begin{equation}}
\def\ee{\end{equation}}
\def\bea{\begin{eqnarray}}
\def\eea{\end{eqnarray}}

\def\met{\mbox{${\hbox{$E$\kern-0.6em\lower-.1ex\hbox{/}}}_T$}} 
\begin{document}
\vspace*{-2.0cm}
\title{Search for Large Extra Dimensions at the Tevatron~\footnote
{To appear in the proceedings of the XXXVI$^{th}$ Rencontres de Moriond, QCD and Hadronic
Interactions, Les Arcs, Savoie, France, March 17-24, 2001.
Full version of this talk is available at {\tt http://moriond.in2p3.fr/}.
}}

\author{ BOB OLIVIER }

\address{Laboratoire de Physique Nucl\'eaire et de Hautes Energies, Universit\'es Paris 6 et 7\\
4 Place Jussieu, Tour 33 RdC, 75252 Paris Cedex 05\\
E-mail: olivier@in2p3.fr}

\maketitle\abstracts{
We report a search for large extra spatial dimensions 
in $p\bar{p}$ collisions at a center of mass energy of 1.8 TeV at the Tevatron.
We present recent D\O\ results on graviton-mediated exchange processes,
using events containing a pair of electrons or photons.
No evidence for signal is found, allowing to place
the most restrictive lower limits on the effective Planck scale at the order of 1 TeV
for several number of extra dimensions.
}

\section{Introduction}
The complete unification of particle
interactions might require the presence of additional spatial dimensions, three
of which are within the reach of our senses, the others being compactified at
distances of the order of $10^{-32}$~m.
In a recent model, inspired by string theory, Arkani-Hamed, Dimopoulos and
Dvali (ADD),$^1$ suggested that there may be only one single scale in particle
phenomena, the electroweak scale, which also corresponds to an
effective Planck scale $M_S \approx 1$~TeV, provided that there exist extra dimensions
compactified at far larger radii than the Planck length, as large as 1~mm.
In this model, particles of the Standard Model (SM) are localized in our
four-dimensional world, but gravitons are allowed to propagate in all the large extra
dimensions. 

Signatures for large extra dimensions depend on whether the gravitons ($G$) in particle
interactions are real (emitted in collisions) or whether they are virtual.
Thus, the impact of virtual gravitons can be 
observed in reactions such as 
$q\bar{q} \rightarrow G \rightarrow \gamma \gamma$, or 
$gg \rightarrow G \rightarrow e^+ e^-$.
Graviton emission can lead to an apparent violation of energy and momentum (as
well as of angular momentum) conservation when the graviton has 
energy momentum components 
transverse to
the brane of the SM, e.g., 
$q\bar{q} \rightarrow G + g$, or
$e^+ e^- \rightarrow G + \gamma$. 
The characteristic signatures for contributions from
virtual graviton correspond to 
abnormally high formation of massive systems, 
while direct emission of graviton results in an increase
of events with large apparent imbalance in transverse momentum
(or ``missing $E_T$'', \met), in particular events with only one jet in the final state
(mono-jet).

Limits on $M_S$ of $\approx 1$~TeV have already been reported from LEP$,^{ 2}$ and
somewhat weaker limits (from searches for virtual graviton contributions) have
been published by experiments at HERA.$^{ 3}$
In this paper we report a search for large extra dimensions at the 
Fermilab Tevatron using approximately 127 $\rm pb^{-1}$
of $p\bar{p}$ data collected at $\sqrt{s}=1.8$~TeV by D\O\ from 1992-96.
\clearpage

\section{Virtual Graviton Effects}
In this paper, we focus on D\O\ results on virtual graviton effects.$^{ 4}$
The amplitude for graviton
exchange has to be added coherently to that of the SM, because processes
such as 
$q\bar{q} \rightarrow Z/\gamma^* \rightarrow e^+ e^- $ and 
$q\bar{q} \rightarrow G  \rightarrow e^+ e^-$ can provide important
interference terms.
Figure 1 represents the Feynman diagrams for dilepton production in the presence of large extra dimensions.

Three phenomenological formulations of the problem have appeared in
the literature.$^{5,6,7}$ They are equivalent, and differ only in their definitions
of $M_S$. D\O\ follows the more sophisticated phenomenology
of Ref.$^5$, which contains
a dependence of the cross section on the number of extra dimensions $n$. 
The correspondence between the
three definitions of scale is that $M_S(n = 5) \approx M_S (\lambda = +1)$ of Ref.$^7$ and
$M_S (n = 4) = \Lambda_T$ of Ref.$^6$.

\begin{figure}
\begin{center}
\vspace*{0.0cm} \epsfig{file=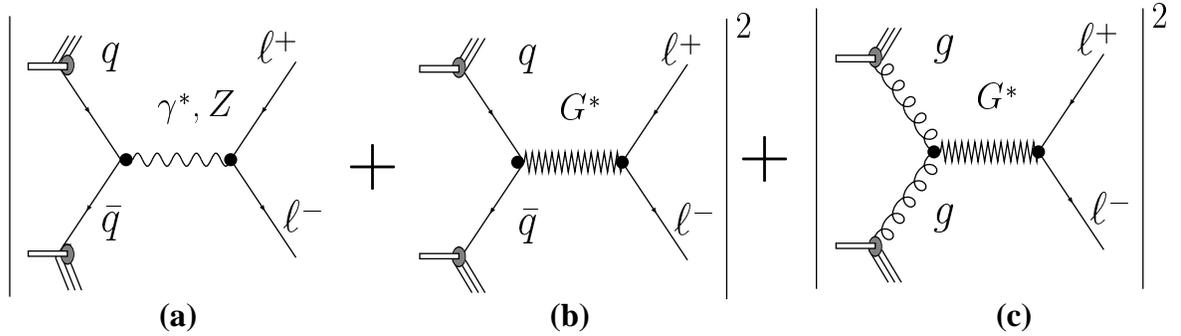,height=4.5cm,bbllx=105pt,bblly=410pt,bburx=554pt,bbury=540pt,clip= } 
\end{center}
\caption{
Feynman diagrams for dilepton production in the presence of large extra dimensions.
$G$ represents graviton exchange. (a) correspond to the SM amplitude contribution, (c) is 
the pure graviton contribution and
(b) the interference between the two.
}
\end{figure}

\section{Analysis at D\O}
D\O\ bases its analysis on both di-electron and di-photon signals. Since D\O\ did
not have a central magnetic field in Run I, the analysis ignores 
the particle charges,
and the cross section is therefore examined as a function of the dielectron
($M_{ee}$) or diphoton ($M_{\gamma\gamma}$) invariant mass,
and $|\cos \theta^*|$, where $\theta^*$ is the angle of the $e$ or $\gamma$ 
relative to the line of flight of the
$ee$ or $\gamma\gamma$ system in the helicity frame. The analysis follows the 
Cheung-Landsberg
extension$^{8}$ of previous studies that were based on the use of only the mass
variable in the problem.

Because the instrumental background (e.g., a jet mimicking a photon or
an electron) in this search is small at large $E_T$,$^{9}$ the signals for $ee$ 
and $\gamma \gamma$
are added together in the comparison of data to theory. This maximizes the
reach of the experiment to highest mass scales because it allows a loosening of
the usual strict electromagnetic (EM) shower requirements on both photons
and electrons. The analysis ignores charged particle tracking information in
the detector, and relies purely on the observation of di-EM systems of high
invariant mass ($M_{EM,EM}$). The theory used to describe the contributions from the
SM and virtual gravitons is a leading order (LO) calculation.
Nevertheless, the
expected yields are corrected for higher order effects through an application of
a common ``K-factor'' of 1.3 (which probably underestimates the expected yield
from gravitons at largest $M_{EM,EM}$, thus resulting in conservative limits 
on their existence). 
The criteria for the final 1250 candidate
events require only two acceptable EM showers, each with $E_T > 45$~GeV, and
no missing transverse momentum ($\met < 25$~GeV). There are no requirements
placed on jets.
\clearpage
\vspace*{-2.0cm}

The data are compared to a LO parton generator,$^{8}$ 
augmented with a parameterized D\O\ detector simulation package 
that models the acceptance, resolution, vertex smearing, impact
of having additional vertices from overlapping multiple interactions,
and introduces a transverse impact
to the di-EM system 
(assuming that it is the same as in the inclusive $Z$ data of D\O$^{10}$). 
As indicated above, the calculation applies a uniform K-factor correction to all SM
and virtual graviton cross sections.
The CTEQ4LO parton
distribution functions (PDF) are used to integrate the matrix elements over
the incident parton distributions (with checks performed using other PDFs).
Most of the di-EM events arise from prompt di-photon, and, to a lesser
extent, from $e^+ e^-$ production (usual SM processes). The background from
other channels, such as $W + $~jets, $W + \gamma$, $WW$, $t\bar{t}$ are negligible. 
The largest instrumental background ($\approx 7\%$ of the di-EM signal) is from multijet
production, where two jets are misidentified as EM showers.

\section{Results}
Figure 2a displays a comparison of data with the Monte Carlo model
containing contributions only from the SM
for the invariant mass of the di-EM system, and shows a good agreement with expectations.$^{8}$

With no excess apparent beyond expectations of the SM, D\O\ proceeds to
calculate a lower limit on the graviton contribution to the di-EM cross section.
The cross section as a function of $M_{EM,EM}$ and 
$|\cos\theta^*|$
can be written as:
\begin{equation}
\sigma = \sigma(SM) + \eta \times \sigma_4 + \eta^2 \times \sigma_8
\end{equation}
where $\eta = F/M_S^4$, with $F = 2/(n-2)$ for $n>$ 2,$^5$ and where $\sigma(SM)$ represents
the SM cross section, $\sigma_8$ the pure graviton contribution, and the term
linear in $\eta$ is the interference between the two. 
The addition of graviton exchange  increases the
yield at large $M_{EM,EM}$, especially for small values of $| \cos \theta^*|$.

The expected sensitivity to $|\eta|$ is obtained from a fit of the above formula 
for the cross section ($\sigma$) to Monte Carlo samples that do not contain
graviton components, and have statistics appropriate to the D\O\ data, which
corresponds to an integrated luminosity of 127 events/pb. Such fits yield an
expected sensitivity of $\eta <$~0.44~$\rm TeV^{-4}$. The D\O\ fits are performed using a
Bayesian formalism that yields the likelihood for $|\eta|$, and the expected limit of
0.44~$\rm TeV^{-4}$
corresponds to an upper limit at 95\% confidence. 

The result of a similar fit to the data yields
an upper limit of  0.46~$\rm TeV^{-4}$,
which provides a lower limit on the value of $M_S$
that depends somewhat on the value of $n$. In particular, $M_S > $~1.44 TeV for
$n$ = 3, and $M_S > $~0.97 TeV for $n$ = 7. 
The results of a similar analysis in the dielectron channel from the CDF collaboration
have recently become available,$^{11}$ and are 
somewhat less restrictive than
the D\O\ ones.
The combined Tevatron limits are expected to yield further improvement 
over the actual excluded range of $M_S$.

\section{Summary}
In summary, D\O\ has presented first results of a search for contributions of
virtual gravitons to production processes at the Tevatron. 
In the context of the ADD scenario
of large extra dimensions with a single mass scale in the domain of particle
interactions,
the D\O\ analysis of massive electrons and photons pairs constrains the 
mass scale $M_S$ to be greater than  1.0$-$1.4 TeV. 
These limits at the 95\% confidence level 
are comparable to the final results anticipated from LEP.
More studies are forthcoming from CDF and D\O\
on real graviton emission (mono-jet events), as well as on virtual graviton
exchange.
By the end of the Tevatron Run II, 
the sensitivity to $M_S$ will reach 3 to 4~TeV.
Beyond that, the LHC will be able to probe effective Planck scales up to 
10~TeV.

\clearpage
\begin{figure}
\vspace*{-1.5cm}
\begin{center}
\begin{tabular}{ll}
\hspace*{-1.0cm}\epsfig{file=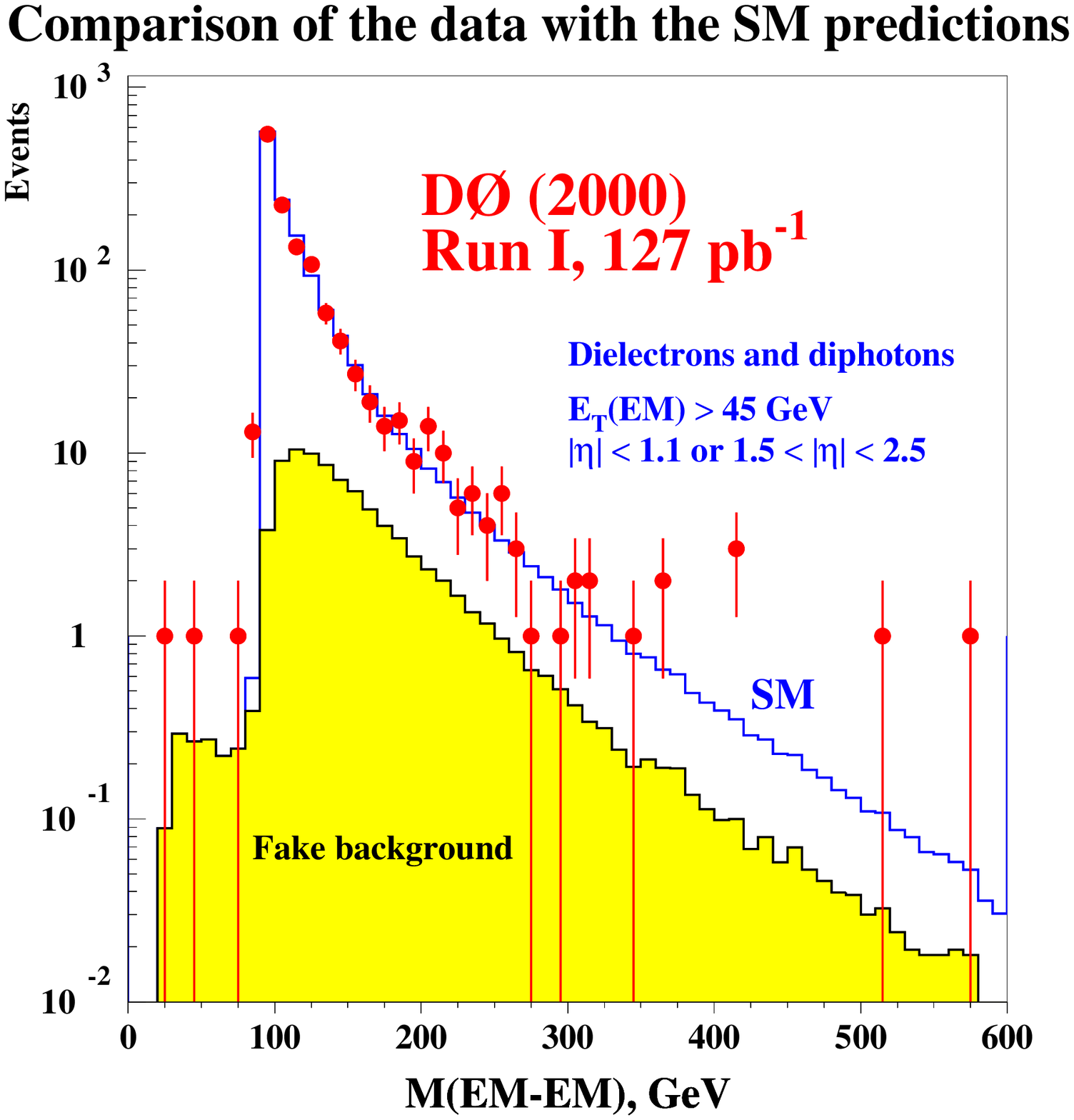,height=8.5cm,bbllx=0pt,bblly=0pt,bburx=567pt,bbury=567pt,clip= }
\epsfig{file=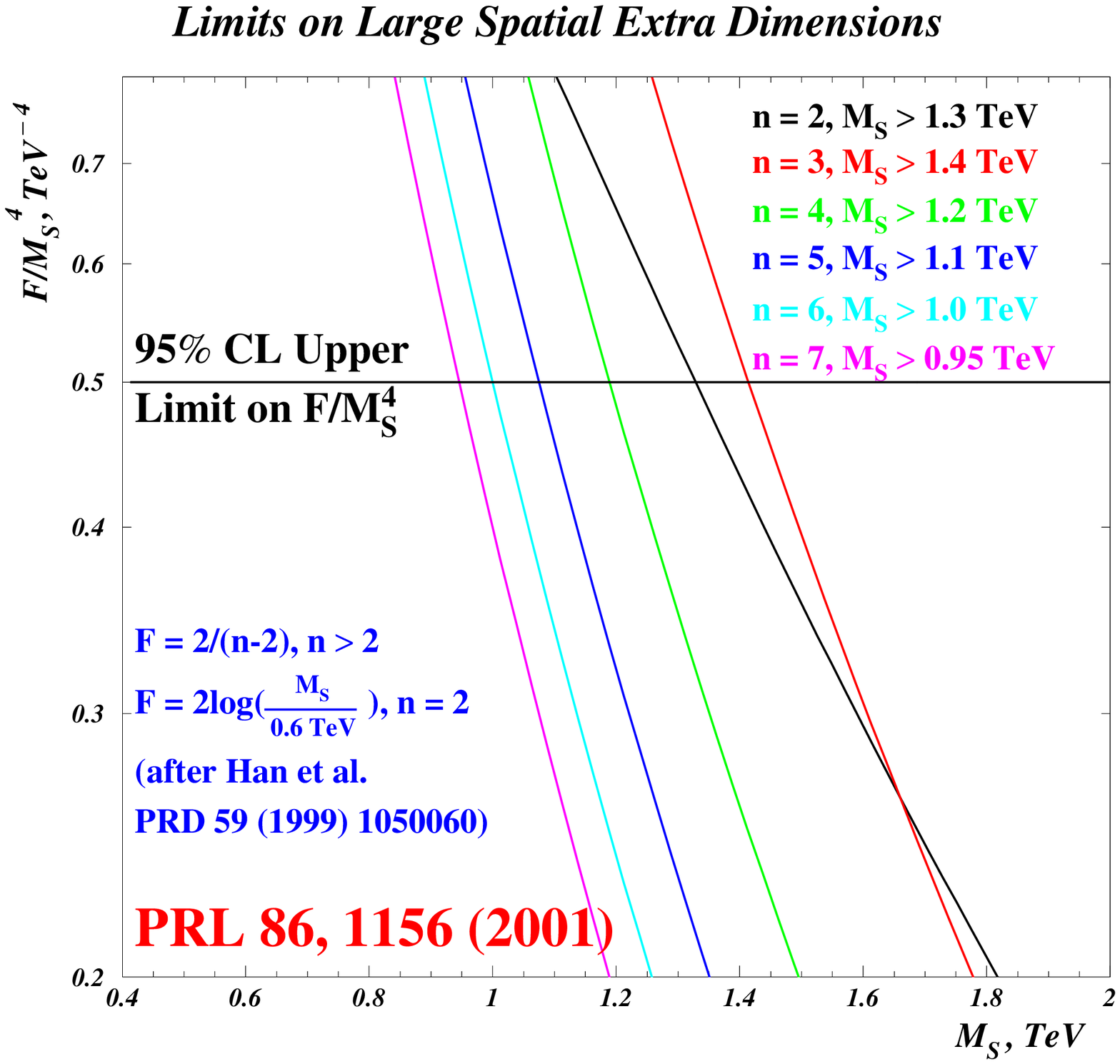,height=8.54cm,bbllx=0pt,bblly=12pt,bburx=567pt,bbury=567pt,clip= } 
\end{tabular}
\begin{picture}(0,0)(-40,0)
\put(-250,190){(a)}
\put(-10,190){(b)}
\end{picture}
\end{center}
\caption{
(a) Comparison of the invariant mass of the di-EM system with the Monte Carlo that contains 
contributions only from the SM; (b) Limits on the effective Planck mass scale $M_S$ 
as a function of the number of extra dimensions.}
\label{3vs4}
\end{figure}

\section*{Acknowledgments}
I thank Greg Landsberg and Tom Ferbel 
for their great help in the preparation of this talk.
I wish to thank the organizers for creating conditions of an exciting conference.

\section*{References}
1. N. Arkani-Hamed, S. Dimopoulos and G. Dvali, Phys. Lett. B {\bf 429}, 263
(1998); I. Antoniadis {\it et al.}, Phys. Lett. B {\bf 436}, 257 (1998); 
N. Arkani-Hamed, S. Dimopoulos and G. Dvali, Phys. Rev. D {\bf 59}, 086004 (1999).\\
2. M. Acciarri {\it et al.}, Phys. Lett. B {\bf 464}, 135 (1999); {\it ibid.}, B {\bf 470}, 268 (1999);
{\it ibid.}, B {\bf 470}, 281 (1999); G. Abbiendi {\it et al.}, Phys. Lett. B {\bf 465}, 303 (1999);
{\it ibid.}, Eur. Phys. J. C {\bf 18}, 253 (2000).\\
3. C. Adloff {\it et al.}, Phys. Lett. B {\bf 479}, 358 (2000).\\
4. B. Abbott {\it et al.}, Phys. Rev. Lett. {\bf 86}, 1156 (2001); See also, 
G. Landsberg, hep-ex/0105039, 
Proc. of XXXVI$^{th}$ Rencontres de Moriond, QCD and Hadronic Interactions, Les Arcs, Savoie, France,
March 17-24, 2001.\\
5. T. Han, J. D. Lykken and R. J. Zhang, Phys. Rev. D {\bf 59}, 105006 (1999).\\
6. G. Giudice, R. Rattazzi and J. Wells, Nucl. Phys. B {\bf 544}, 3 (1999).\\
7. J. L. Hewett, Phys. Rev. Lett. {\bf 82}, 4765 (1999).\\
8. K. Cheung and G. Landsberg, Phys. Rev. D {\bf 62}, 076003 (2000).\\
9. G. Landsberg and K. T. Matchev, Phys. Rev. D {\bf 62}, 035004 (2000).\\
10. B. Abbott {\it et al.}, Phys. Rev. Lett. {\bf 84}, 2792 (2000).\\
11. J. Carlson, talk at the APS meeting, Washington, D.C., April 28-May 1, 2001.

\end{document}